# 3D Conductive Polymer Printed Metasurface Antenna for Fresnel Focusing

Okan Yurduseven, *Senior Member, IEEE*, Shengrong Ye, Thomas Fromenteze, Daniel L. Marks, Benjamin J. Wiley, and David R. Smith, *Senior Member, IEEE*

*Abstract*—**We demonstrate a 3D printed holographic metasurface antenna for beam-focusing applications at 10 GHz within the X-band frequency regime. The metasurface antenna is printed using a dual-material 3D printer leveraging a biodegradable conductive polymer material (Electrifi) to print the conductive parts and polylactic acid (PLA) to print the dielectric substrate. The entire metasurface antenna is 3D printed at once; no additional techniques, such as metal-plating and laser etching, are required. It is demonstrated that using the 3D printed conductive polymer metasurface antenna, high-fidelity beam focusing can be achieved within the Fresnel region of the antenna. It is also shown that the material conductivity for 3D printing has a substantial effect on the radiation characteristics of the metasurface antenna.**

*Index Terms*—**Additive manufacturing, 3D printing, conductive polymer, microwaves, metamaterials, metasurface, beam-focusing, near-field, Fresnel.**

## I. INTRODUCTION

ANTENNAS with electromagnetic (EM) wave-front control capabilities offer unique opportunities with numerous applications in near-field (or Fresnel) [1-3] and far-field (or Fraunhofer) [4-6] beam-forming. The concept of Fresnel focusing enables the antenna radiated fields to be focused at a defined spot in the Fresnel region of the antenna, leading to increased field intensities at the focused region. This capability has been shown to hold significant potential in several emerging applications, such as near-field imaging [7], non-destructive testing [8], and wireless power transfer [9].

Conventionally, these antennas are realized using printed-circuit-board (PCB) technology. In microwave and millimeter-wave frequency regimes, conventional fabrication techniques include photolithography or laser-based etching. Although high accuracy at extremely fine resolution limits can be achieved, these techniques can be limited in a number of key system metrics, such as exhibiting a rather complex fabrication process and long prototyping times are required. Recently, additive manufacturing (or 3D printing) of EM designs and antennas has gained traction as a means of offering rapid prototyping of EM structures [10-18].

In this paper, we demonstrate a 3D printed metasurface antenna for Fresnel focusing. The metasurface layer of the antenna is made of a biodegradable conductive polymer material, Electrifi, printed on a polylactic acid (PLA) dielectric substrate. The conductive Electrifi material can easily be printed using the Fused Deposition Modeling (FDM) technology. As opposed to conventional EM designs and 3D printing methods, we demonstrate that the proposed 3D printing technique enables the antenna to be fabricated at once, circumventing the requirement for metal plating techniques, which can pose difficulties in plating complex structures. As the 3D printed conductive and dielectric layers of the proposed antenna are fully-integrated during the printing process, no additional assembly methods are needed.

## II. 3D PRINTED METASURFACE ANTENNA DESIGN

The 3D printed Fresnel focusing metasurface antenna is depicted in Fig. 1. As shown in Fig. 1, the design consists of a PLA substrate ($\varepsilon_r$=3 and tan$\delta$=0.02) sandwiched in between the metasurface layer on top and a ground plane at the bottom, both of which are printed using Electrifi. The metasurface layer is patterned into an array of subwavelength slot-shaped metamaterial elements (or meta-elements). These meta-elements couple to the guided-mode (or the reference-wave) launched into the PLA substrate by a coaxial feed placed in the center of the antenna. The meta-elements are $\lambda_g/3$ long and $\lambda_g/10$ wide, where $\lambda_g$ is the guided wavelength within the dielectric at 10 GHz, 1.73 cm. The aperture of the antenna is discretized into a regular grid of square pixels of $\lambda_g/2.5$ size, corresponding to $\lambda_0/4.3$ in free-space, smaller than the periodicity of radiating elements in conventional array antennas ($\lambda_0/2$).

The PLA substrate is $t_s$=3 mm thick, forming a 3D volumetric structure while ensuring the rigidity of the antenna and enabling

This work was supported by the Air Force Office of Scientific Research (AFOSR, Grant No. FA9550-12-1-0491).

O. Yurduseven was with the Department of Electrical and Computer Engineering, Duke University, Durham, North Carolina 27708 USA. He is now with the Jet Propulsion Laboratory, California Institute of Technology, Pasadena, CA 91106 USA (e-mail: okan.yurduseven@jpl.nasa.gov).

S. Ye is with Multi3D LLC, 101-U Woodwinds Industrial Court. Cary, North Carolina 27511 USA.

T. Fromenteze is with the Xlim Research Institute, University of Limoges, 87060, Limoges, France.

D. L. Marks and D. R. Smith are with the Department of Electrical and Computer Engineering, Duke University, Durham, North Carolina 27708 USA.

B. J. Wiley is with the Department of Chemistry, Duke University, North Carolina 27708 USA.

single mode operation, $t<\lambda_g/2$. The Electrifi metasurface layer and the ground plane are $t_e=0.4$ mm thick, larger than the skin depth of the material at 10 GHz, $\delta=38.9$ μm.

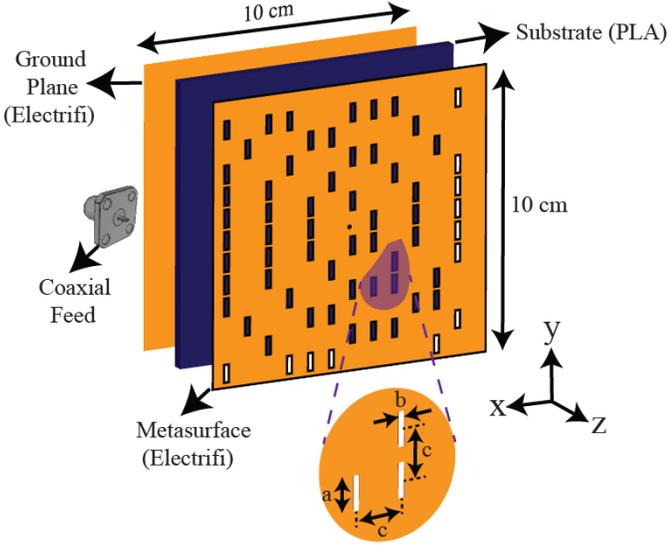

**Fig. 1.** Depiction of a 3D printed holographic metasurface antenna for Fresnel focusing. Layers are shown separated for illustration. Meta-element dimensions; $a=\lambda_g/3$, $b=\lambda_g/10$, $c=\lambda_g/2.5$ where guided-wavelength, $\lambda_g$, is 1.73 cm.

Focusing at an arbitrarily selected point in the Fresnel region requires that the wavefront within each discretized pixel on the aperture adds up constructively at the focal point. Therefore, to achieve Fresnel focusing, the guided-mode reference is modulated by the metasurface layer to result in the desired field distribution on the antenna aperture that would produce the focus of interest [5]. This suggests that the metasurface layer acts as a holographic modulator, producing the focused beam when illuminated by the guided-mode reference-wave. In this work, we make use of a binary modulation scheme, involving treating the focal point as a virtual point source back-propagated to the aperture of the antenna, and placing the meta-elements only at the points where the phase difference between the back-propagated aperture field and the guided mode reference remains below a certain threshold. In this work, the phase threshold was found optimum to be ±60⁰. Increasing the phase threshold beyond this level increases the number of meta-elements in the metasurface layer, resulting in a stronger radiation from the antenna at the expense of increasing the sidelobe levels (worsening constructive interference). Reducing this phase threshold, on the other hand, reduces the number of meta-elements, and therefore the radiation from the antenna.

In Fig. 2, we depict the holographic beam-focusing concept using the 3D printed metasurface antenna. The first step in this process consists of determining the virtual point source at the position where beam-focusing is desired to be achieved with the position of the point source being **r'** in Fig. 2. The virtual point source is then back-propagated to the aperture of the antenna, $\mathbf{P}=e^{-jk|\mathbf{r}-\mathbf{r}'|}$ (amplitude dependency is dropped as beam fidelity is mainly controlled by phase information), producing a desired field distribution, **P**, that would focus at the position of the virtual point source when radiated by the metasurface antenna with the coordinates of the antenna being **r** in Fig. 2. The second step in the design process is to design a metasurface layer that would produce **P** when excited by the magnetic field reference-wave launched by the coaxial feed, $\mathbf{H}=H_0^{(1)}(k_g\mathbf{r})\sin\phi$. Here, $k_g$ is the wavenumber within the substrate, $H_0^{(1)}$ is the Hankel function (zeroth order and first kind) and we consider only the y-component of the launched magnetic field exciting the slots with their longitudinal axes oriented along the y-axis. From this definition, the metasurface can be considered a phase grating, $\mathbf{M}=\mathbf{PH}^\dagger$ where symbol † denotes the complex conjugate operation. It should be noted that the fields are scalar and we use the bold font to denote the vector-matrix notation.

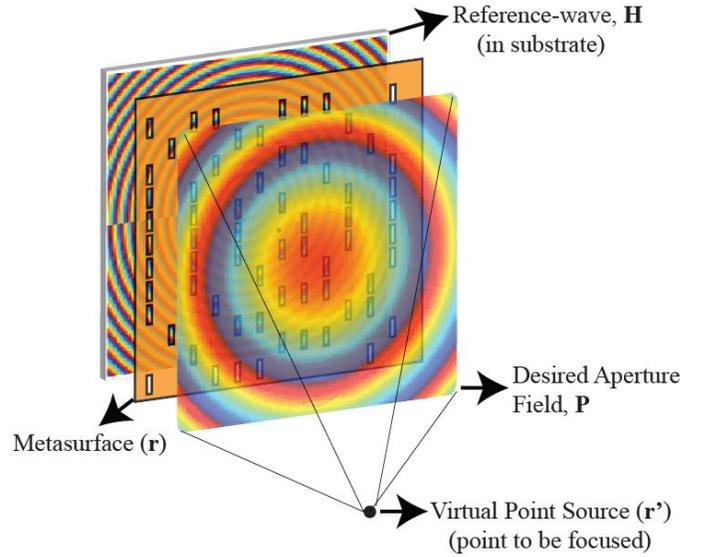

**Fig. 2.** Depiction of the holographic beam-forming process.

The Electrifi conductive filament is a copper-polyester composite filament which has a conductivity of $\sigma=1.67\times10^4$ S/m. The design of our 3D printer is based on an open source FDM D-Bot printer [19]. The original design was converted from a single Bowden extruder to a dual direct drive extruder. As a result, the 3D printer is capable of printing both the Electrifi and the PLA materials simultaneously. The nozzle size for printing is 0.5 mm and the print speed is 15 mm/s for Electrifi and 30 mm/s for PLA. The printing speed for the Electrifi material is 140 ⁰C while the PLA material is printed at 190 ⁰C. To print the Electrifi material, no heated bed is used, enabling us to maintain the maximum conductivity of the Electrifi material.

In this paper, we investigate Fresnel focusing for two different scenarios; on-axis and off-axis. From antenna theory [20], the far-field limit of the designed aperture is calculated as $z_{max}=0.66$ m, with the focal points selected to be $F_1(x=0$ m, $y=0$ m, $z=0.2$ m) and $F_2(x=-0.05$ m, $y=0.05$ m, $z=0.2$ m), respectively. Here, $F_1$ denotes on-axis focusing scenario where the focusing is achieved along the broadside direction of the antenna (z-axis) while $F_2$ denotes the off-axis focusing scenario. The 3D printed metasurface antennas for the investigated focusing scenarios are shown in Fig. 3.

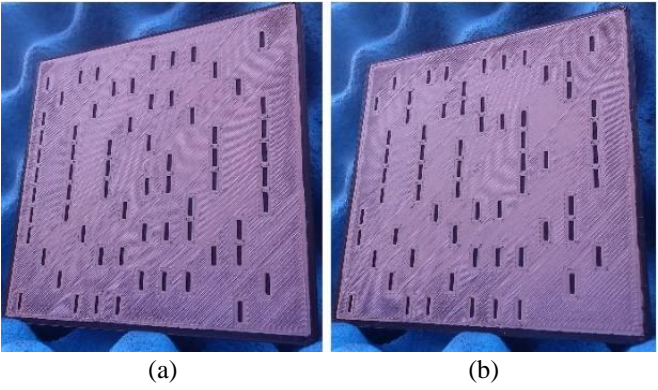

(a)    (b)

**Fig. 3.** 3D printed conductive polymer metasurface antennas for on-axis and off-axis Fresnel focusing (a) on-axis focusing, F₁(x=0 m, y=0 m, z=0.2 m) (b) off-axis focusing, F₂(x=-0.05 m, y=0.05 m, z=0.2 m).

As shown in Fig. 3, the distribution of the meta-elements in the metasurface layer changes as the focal point is varied. This is expected as the fields back-propagated from F₁ and F₂ to the antenna aperture are different, producing a different hologram pattern when interacted with the guided-mode reference. The surface roughness of the finished prototypes shown in Fig. 3 were measured using a surface profilometer, Bruker Dektak 150 [21] and the recorded to be 22 μm.

### III. RESULTS AND DISCUSSION

To characterize the fields radiated by the antennas at the focal distance, we make use of a near-field scanning system (NSI-200v-3x3). The measured electric field (E-field) patterns at $z=0.2$ m plane are shown in Fig. 4. In Fig. 4 the on-axis and off-axis focused beams are evident. It should be noted that for this demonstration F₁ and F₂ focal points are chosen on an arbitrary basis and the antenna can focus at another desired point of interest within the Fresnel zone of the aperture.

An important parameter to assess the focusing capabilities of the 3D printed metasurface antennas is the beam-waist diameter of the E-field patterns at the focal plane, $z=0.2$ m. Analyzing the measured E-field patterns shown in Fig. 4, the -3 dB full-width-half-maximum (FWHM) beam-waist values are calculated to be 6.82 cm for on-axis and 7.48 cm for off-axis focusing scenarios, respectively. These values are in good agreement with the theoretical beam-waist values which are calculated to be 6.48 cm for on-axis and 6.89 cm for off-axis focusing scenarios, respectively [9]. We note that for the calculation of the theoretical beam-waist values, the Gaussian optics limits presented in [9], which conventionally take $1/e^2$ width level as a reference, were converted to a FWHM scale, which is more common in antenna community.

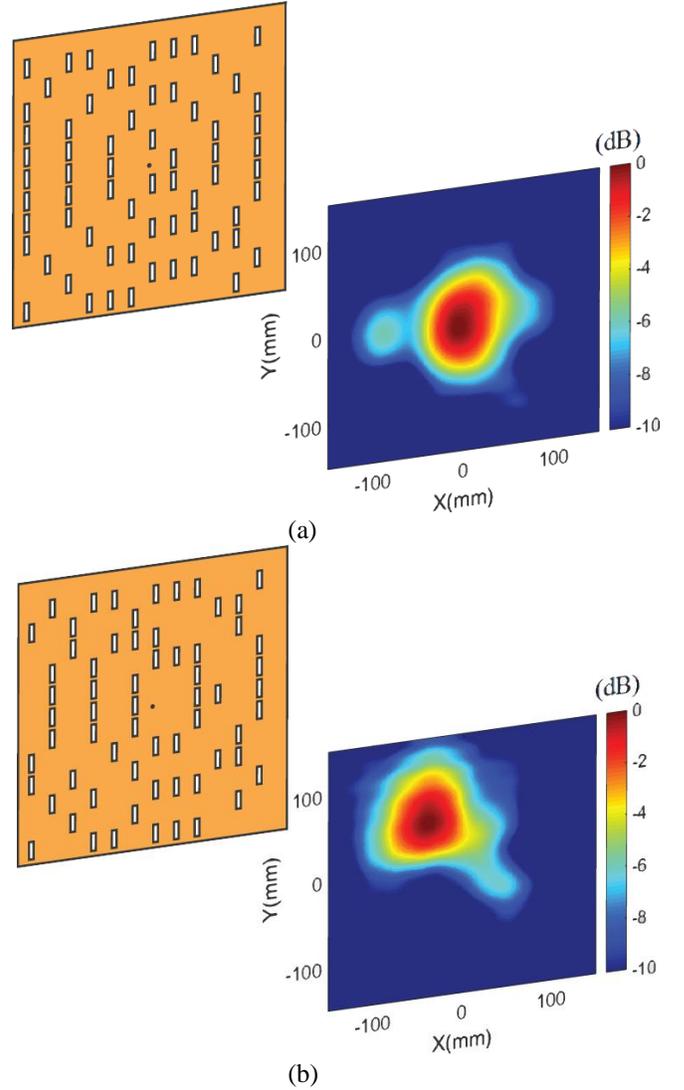

(a)

(b)

**Fig. 4.** Measured E-field patterns at the focal plane, $z=0.2$ m (a) on-axis focusing (b) off-axis focusing.

Conventionally, an antenna is designed for far-field operation. Analyzing antennas in the far-field is a well-defined problem with certain metrics used for performance evaluation, including antenna gain, sidelobe level and -3 dB half-power-beamwidth (HPBW). In the near-field, however, the radiation pattern cannot be considered in its conventional far-field sense due to the presence of angular-dependency with distance. However, to analyze the effect of the material conductivity on the antenna performance, we can use the relative change in these parameters as a function of varying material conductivity. To this end, we vary the conductivity of the Electrifi material within the range of $\sigma_{min}=1.67\text{x}10^3$ and $\sigma_{max}=1.67\text{x}10^5$, and analyze the conductivity-dependency of the gain, sidelobe level and HPBW values of the metasurface antenna in Table I. This analysis was performed using a full-wave EM simulator, CST Microwave Studio, with lossy (tanδ=0.02) and lossless (tanδ=0) PLA material as the substrate.

**Table I.** Antenna performance as a function of material conductivity and dielectric loss.

| PLA Substrate Loss (tan$\delta$) | Material Conductivity (S/m) | Gain (dBi) | Sidelobe Level (dB) | HPBW (Degrees) |
|---|---|---|---|---|
| N/A | 1.67x10$^3$ | 1.33 | -11.4 | 22.4 |
| | 1.67x10$^4$ | 7.29 | -13.2 | 20.4 |
| | 1.67x10$^5$ | 11 | -13.6 | 18.9 |
| 0.02 | 1.67x10$^3$ | 0.34 | -11.3 | 23.2 |
| | 1.67x10$^4$ | 5.31 | -13.1 | 21.1 |
| | 1.67x10$^5$ | 7.55 | -15 | 20.1 |

Several important conclusions can be drawn from Table I. First, increasing the Electrifi material conductivity significantly increases the gain of the antenna. As an example, increasing the material conductivity from $\sigma$=1.67x10$^3$ to $\sigma$=1.67x10$^5$, the gain is increased by 9.67 dB and 7.21 dB for the lossless and lossy PLA substrate scenarios, respectively. Second, albeit being limited in comparison to the effect of the material conductivity, the dielectric loss of the PLA material has a considerable effect on the gain performance of the antenna. Taking the actual conductivity of the Electrifi material as a reference, $\sigma$=1.67x10$^4$ S/m, the dielectric loss of the PLA substrate is responsible for a 1.98 dB difference in the gain value. Third, increasing the dielectric loss of the substrate reduces the sidelobe levels while widening the HPBW. This is expected as increasing the loss of the PLA substrate has a similar effect to truncating the aperture, resulting in a smaller effective aperture size. Fourth, and finally, as the material conductivity is increased, the sidelobe levels improve while the HPBW becomes narrower.

## IV. Conclusion

We have demonstrated a 3D printed conductive polymer metasurface antenna for Fresnel focusing applications. Using the proposed antenna, high-fidelity beam focusing at arbitrarily selected focal points in the near-field zone of the antenna has been achieved. It has also been observed that improving the material conductivity can significantly enhance the radiation characteristics of the proposed antenna. While the 3D printing biodegradable Electrifi material used in this design has an electrical conductivity of $\sigma$=1.67x10$^4$, our ongoing studies suggest that a conductivity increase of 10 times can be achieved by increasing the amount of the copper in Electirifi. The proposed conductive polymer metasurface antenna is low-cost, simple to manufacture, and suitable for rapid prototyping. It can find applications in near-field imaging, non-destructive testing and wireless power transfer, where rapid prototyping can be a significant advantage in the design process.

## References

[1] M. Ettorre, *et al.*, "On the Near-Field Shaping and Focusing Capability of a Radial Line Slot Array," *IEEE Transactions on Antennas and Propagation*, vol. 62, no. 4, pp. 1991-1999, 2014.

[2] A. J. Martinez-Ros, J. L. Gómez-Tornero, V. Losada, F. Mesa and F. Medina, "Non-Uniform Sinusoidally Modulated Half-Mode Leaky-Wave Lines for Near-Field Focusing Pattern Synthesis," in *IEEE Transactions on Antennas and Propagation*, vol. 63, no. 3, pp. 1022-1031, 2015.

[3] D. Blanco, J. L. Gómez-Tornero, E. Rajo-Iglesias and N. Llombart, "Radially Polarized Annular-Slot Leaky-Wave Antenna for Three-Dimensional Near-Field Microwave Focusing," *IEEE Antennas and Wireless Propagation Letters*, vol. 13, pp. 583-586, 2014.

[4] A. Epstein, and G. V. Eleftheriades, "Passive Lossless Huygens Metasurfaces for Conversion of Arbitrary Source Field to Directive Radiation," *IEEE Transactions on Antennas and Propagation*, vol. 62, no. 11, pp. 5680-5695, 2014.

[5] O. Yurduseven, D. L. Marks, T. Fromenteze, and D. R. Smith, "Dynamically reconfigurable holographic metasurface aperture for a Mills-Cross monochromatic microwave camera," *Optics Express*, vol. 26, no. 5, pp.5281-5291, 2018.

[6] M. C. Johnson, S. L. Brunton, N. B. Kundtz and J. N. Kutz, "Sidelobe Canceling for Reconfigurable Holographic Metamaterial Antenna," in *IEEE Transactions on Antennas and Propagation*, vol. 63, no. 4, pp. 1881-1886, 2015.

[7] O. Yurduseven, D. L. Marks, T. Fromenteze, J. N. Gollub, and D. R. Smith, "Millimeter-wave spotlight imager using dynamic holographic metasurface antennas," *Optics Express*, vol. 25, no. 15, pp. 18230-18249, 2017.

[8] C. Ziehm, S. Hantscher, J. Hinken, C. Ziep, and M. Richter"Near field focusing for nondestructive microwave testing at 24 GHz–Theory and experimental verification," *Case Studies in Nondestructive Testing and Evaluation*, vol. 6, pp. 70-78, 2016.

[9] D. R. Smith, et al., "An analysis of beamed wireless power transfer in the Fresnel zone using a dynamic, metasurface aperture," *Journal of Applied Physics*, vol. 121, no. 1, pp. 014901, 2017.

[10] M. Liang, C. Shemelya, E. MacDonald, et al., "3-D Printed Microwave Patch Antenna via Fused Deposition Method and Ultrasonic Wire Mesh Embedding Technique," IEEE Antennas and Wireless Propagation Letters, vol. 14, pp. 1346-1349, 2015.

[11] B. T. W. Gillatt, M. D'Auria, W. J. Otter, "3-D Printed Variable Phase Shifter," IEEE Microwave and Wireless Components Letters, vol. 26, no. 10, pp. 822-824, 2016.

[12] A. T. Castro, B. Babakhani and S. K. Sharma, "Design and development of a multimode waveguide corrugated horn antenna using 3D printing technology and its comparison with aluminium-based prototype," in *IET Microwaves, Antennas & Propagation*, vol. 11, no. 14, pp. 1977-1984, 11 19 2017.

[13] S. Zhang, "Design and fabrication of 3D-printed planar Fresnel zone plate lens," *Electronics Letters*, vol. 52, no. 10, pp. 833-835, 2016.

[14] H. Xin and M. Liang, "3-D-Printed Microwave and THz Devices Using Polymer Jetting Techniques," *Proceedings of the IEEE*, vol. 105, no. 4, pp. 737-755, 2017.

[15] O. Yurduseven, et al., "Computational microwave imaging using 3D printed conductive polymer frequency-diverse metasurface antennas," *IET Microwaves, Antennas & Propagation*, vol. 11, no. 14, pp. 1962-1969, 2017.

[16] I. M. Ehrenberg, S. E. Sarma, and B. I. Wu, "A three-dimensional self-supporting low loss microwave lens with a negative refractive index," *Journal of Applied Physics*, vol. 112, no. 7, p.073114, 2012.

[17] H. Yi, S. W. Qu, K. B. Ng, C. H. Chan and X. Bai, "3-D Printed Millimeter-Wave and Terahertz Lenses with Fixed and Frequency Scanned Beam," *IEEE Transactions on Antennas and Propagation*, vol. 64, no. 2, pp. 442-449, Feb. 2016.

[18] S. S. Bukhari, W. G. Whittow, S. Zhang and J. C. Vardaxoglou, "Composite materials for microwave devices using additive manufacturing," *Electronics Letters*, vol. 52, no. 10, pp. 832-833, 5 12 2016.

[19] 'D-Bot Core-XY 3D Printer', http://www.thingiverse.com/thing:1001065, accessed 5 March 2018.

[20] C. A. Balanis, *Antenna theory: analysis and design*, John Wiley & Sons, 3$^{rd}$ edition, 2005.

[21] 'Bruker Dektak 150', https://www.brukersupport.com/ProductDetail/1135, accessed 10 March 2018.